\documentstyle[aps,multicol,epsf]{revtex}
\input epsf
\begin{document}
\draft

\title{Numerical Study of a Superconducting Glass Model}
\author{J.M. Kosterlitz and M.V. Simkin}
\address{Dept. Physics, Brown University,\\
	  Providence, RI 02912-1843}
\maketitle

\widetext

\begin{abstract}
An $XY$ model with random phase shifts as a model for a superconducting
glass is studied in two and three dimensions by a zero temperature domain
wall renormalization group which allows one to follow the flows of both the
coupling constant and the disorder strength with increasing length scale.
Weak disorder is found to be marginal in two and probably irrelevant in three
dimensions. For strong disorder the flow is towards a non-superconducting
gauge glass fixed point in $2d$ and a superconducting glass in $3d$. Our
results are in agreement with recent analytic theory and are inconsistent
with earlier predictions of a re-entrant transition to a disordered phase at
very low temperature and with the loss of superconductivity for any finite
amount of disorder.
\end{abstract}

\begin{multicols}{2}
\narrowtext

 The classical $XY$ model with random phase shifts described by the
Hamiltonian\cite{es}
\begin{equation}
H=-\sum_{<ij>}J_{ij}cos(\theta_{i}-\theta_{j}-A_{ij})
\label{1}
\end{equation}
has been the subject of much interest over the past decade. Here,
$\theta_{i}$ is the phase of the order parameter at
the $i^{th}$ site of a square lattice in $2d$ or a simple cubic
lattice in $3d$ and the sum is over all nearest neighbor bonds. The coupling
constants $J_{ij}$ are uniform $J_{ij}=J>0$ and the quenched random phase
shifts $A_{ij}$ are uncorrelated from bond to bond and uniformly distributed
over the range $-\alpha\pi\leq A_{ij}\leq\alpha\pi$ with $0\leq\alpha\leq 1$
so that $<A_{ij}>=0$ and $<|A_{ij}|>=\alpha\pi /2$ where $<>$ means an
average over disorder. In $d=2$, the model
of eq.(\ref{1}) is a description of an array of Josephson junctions in a
magnetic field perpendicular to the plane of the array when $\theta_{i}$
represents the phase of the superconducting order parameter of the $i^{th}$
grain and $A_{ij}=(2\pi/\Phi_{0})\int_{i}^{j}\vec A\cdot d\vec l$ with $\vec
A$ the vector potential of the external magnetic field and $\Phi_{0}=hc/2e$
is the quantum of flux. The $A_{ij}$ become independent quenched random
variables when the average flux through an elementary plaquette is an
integer multiple of $\Phi_{0}$ but the superconducting grains are randomly
displaced from their ideal lattice positions\cite{gk}.
It is also a model for
an $XY$ magnet with random Dzyaloshinskii-Moriya interactions\cite{rsn}.

Whatever the physical origin of the quenched random phase shifts, the system
described by eq.(\ref{1}) has been a theoretical challenge for a decade.
Early work concluded that weak disorder ($\alpha\ll 1$) does not destroy the
superconducting phase at intermediate temperature $T$ but at low $T$ there
is a re-entrant transition to a normal phase\cite{rsn,gk}. However, this was
not confirmed either numerically nor experimentally\cite{exp}. Some later
theoretical work\cite{kor} suggested that the bound vortex (KT)
phase\cite{kt}
is destroyed at any finite $T$ by arbitrarily small disorder and that the
experimentally observed KT phase is a finite size effect. The more recent
theoretical work\cite{nskl,kn,scheidl,tang} based on the ideas of Cha and
Fertig\cite{cha}, on the other hand, argue for a more conventional phase
diagram in which there is a superconducting phase for $T<T_{c}(\alpha)$
where $T_{c}(\alpha)\geq 0$ for $\alpha\leq\alpha_{c}$. The case of
maximum disorder
($\alpha =1$) when the $A_{ij}$ are uniformly distributed beween $-\pi$
and $+\pi$ is called the gauge glass\cite{mpaf,hs} which has been studied
numerically at
$T=0$ by a domain wall renormalization group (DWRG)\cite{mjpg,fty,rtyf}
in both $d=2$
and $d=3$ and the conclusions from these studies are that the stiffness to
distortions of the phase vanishes in $d=2$ so that the glass is not
superconducting and that in $d=3$ the gauge glass is probably
superconducting, but the evidence is not conclusive. In view of the three
conflicting scenarios in two dimensions, (i) re-entrant
transition\cite{rsn,gk}, (ii) destruction of superconductivity
for any finite disorder\cite{kor} and (iii) superconductivity for
$T<T_{c}(\alpha)$ with $T_{c}(\alpha)>0$ for
$0\leq\alpha<\alpha_{c}$\cite{nskl,scheidl,tang}, this
system is an ideal candidate for study by a numerical DWRG at $T=0$ as this
will distinguish scenario (iii) from the others as only scenario (iii)
predicts a finite stiffness at $T=0$ for $0\leq\alpha <\alpha_{c}$.

The standard DWRG\cite{dwrg} at $T=0$ consists of computing the lowest
energies of
a set of systems of several linear sizes $L$ with periodic and antiperiodic
boundary conditions (BC) in one direction
with some fixed BC in the other $d-1$ directions. The difference $\Delta
E(L)=<|E_{ap}(L)-E_{p}(L)|>$ is the domain wall energy and $\Delta E(L)/2$
is interpreted as an effective coupling constant $J(L)$ at length scale $L$
which one expects to scale as $J(L)\sim L^{\theta}$ at large $L$. The
stiffness exponent $\theta$ is a crucial quantity as its value will
distinguish between an ordered superconducting phase at small but finite $T$
($\theta\geq 0$) and a disordered phase ($\theta <0$). If $\theta <0$, then
the energy of an excitation of size $L$ is $\Delta E(L)$ which vanishes as
$L\rightarrow\infty$ and the probability of such a phase unwinding
excitation $P(\Delta E(L))\sim exp(-\Delta E(L)/kT)$ is large at any $T>0$
so the stiffness to twists in the phase will vanish. If $\theta >0$, the
converse is true and the stiffness will be finite for $T<T_{c}$ and the
glass will be superconducting. This technique has been applied
to random systems such as spin and gauge glasses at
$T=0$ but the the value of the exponent $\theta$ obtained by this method
is not very reliable because it is not clear if the asymptotic scaling regime
is reached for the small sizes $L$ it is usually possible to simulate.

For the problem of interest with a variable disorder strength, this
version of the DWRG, which considers only the scaling of the effective
coupling $J(L)$ needs considerable modification as we also want to know how
the
disorder strength scales with $L$. For a single junction, the Hamiltonian is
$H=-Jcos(\phi -A)$ where $\phi$ is the phase difference across the junction
and the usual comparision of the energies with periodic and antiperiodic BC
gives $\Delta E(1)=2JcosA$ which does not separate the disorder strength
from the coupling constant. At length scale $L$, the interaction is
$V_{L}(\phi -A(L))$ where $A(L)$ is the phase shift at scale $L$ and
$V_{L}(\phi)$ is a $2\pi$-periodic function with a minimum when its argument
is zero. Thus, if we impose BC with phase shifts $\Delta_{\mu}$ across the
boundaries in the $d$ directions $\mu =1,2,...d$, minimizing the energy
with
respect to the phases $\theta_{i}$ will give the ground state energy of
a system of linear size $L$ as a function of $\Delta_{\mu}$ which is
$2\pi$ periodic in each of the $d$ directions
\begin{equation}
E_{L}(\Delta_{\mu})=E_{L}(\Delta_{\mu}+2\pi )
\end{equation}
with a minimum at some $\Delta_{\mu}^{0}$ which depends on the precise
realization of disorder. The key observation is that $\Delta_{\mu}^{0}$ is
exactly the phase shift $A_{\mu}(L)$ which minimizes the energy at scale
$L$. A measure of the strength of disorder at this scale is
\begin{equation}
|A(L)|\equiv <|\Delta^{0}|>
\end{equation}
with $|A(1)|=\alpha\pi/2$. The coupling constant $J(L)$ at scale $L$ is
found by first finding
$E_{L}(\Delta_{\mu}^{0})$, changing $\Delta^{0}$ by $\pi$ in one
of the $d$ directions and then finding the energy minimum
$E_{L}(\Delta^{0}+\pi)$ with these BC. As discussed above, the coupling
constant $J(L)$ at scale $L$ is
\begin{equation}
J(L)\equiv <(E_{L}(\Delta^{0}+\pi)-E_{L}(\Delta^{0}))>
\end{equation}
and measuring $J(L)$ and $|A(L)|$ for several sizes $L$ gives
renormalization
group flows for both the coupling constant and disorder strength.
Of interest are the stable fixed point values $J^{*}\equiv J(L=\infty)$ and
$A^{*}\equiv |A(L=\infty)|$ as these determine the nature of the phases.
There are several possibilities of which the simplest are $[J^{*}=\infty$,
$A^{*}=0]$, $[J^{*}=\infty$,  $A^{*}=\pi/2]$, $[J^{*}=0$, $A^{*}=\pi/2]$
corresponding respectively to a superconducting state with long range order,
a supeconducting glass and a non-superconducting glass. There are other
possibilities such as a state with quasi long range order corresponding to
a flow to a fixed line with finite $J^{*}$ and $A^{*}$ whose values depend
on the initial values of coupling and disorder. This is the scenario in
$d=2$ predicted by recent analytic work\cite{scheidl,tang}.

Of course, since we do not know how to find the exact ground state of a
disordered system of arbitrary linear size $L$, the best we can do is to
numerically estimate $J(L)$ and $|A(L)|$ for a set of samples of different
sizes $L$ up to some maximum and extrapolate to large $L$. We use simulated
annealing\cite{kgv} to estimate the ground state energies which is
considerably
more efficient than simple repeated quenches to $T=0$\cite{sim}. Also, we
imposed periodic $\Delta_{\mu} =0$ BC in $d-1$
directions and twisted $\Delta\neq 0$ BC in the remaining direction
and minimized the energy with respect to the the phases $\theta_{i}$
and to the twist $\Delta$ to find $\Delta^{0}$. To obtain the domain wall
energy $\Delta E_{L}$ the twist is changed to $\Delta^{0}+\pi$ and kept
fixed while the energy is minimized with respect to the $\theta_{i}$ only.
According to our earlier discussion, the energy should be
minimized with respect to global phase shifts in all $d$ directions and the
domain wall energy $\Delta E_{L}$ obtained by increasing the phase shift by
$\pi$ in one direction. To within the errors of our simulations, $\Delta
E_{L}$ is independent of the choice of BC in the $d-1$ transverse directions
so, for simplicity, we imposed periodic or $\Delta_{\mu}=0$ BC in these
directions.
As a consistency check\cite{sim}, we simulated two identical copies of each
system
with different random number sequences to obtain two estimates $E_{1},E_{2}$
of the ground state energy. In the event that the simulation finds the exact
ground state, then $\delta E=E_{1}-E_{2}=0$, which often ocurrs for our
small $L$ values. If the simulation does not reach the exact minima,
$<(\delta E)^{2}>$ is a measure of the error.
To minimize the errors caused by failure to reach the true energy
minimum, we adjust the annealing schedule and the number of annealing
attempts until $\delta E/E<N^{-1/2}$ where $N=10^{3}$ in $2d$ and $10^{4}$
in $3d$ is the number of realizations
of disorder. This consistency check makes the error due to not reaching the
true ground state no worse than the statistical error in the averaging over
disorder. To our knowledge, there is no analogue for this disordered $XY$
system of the "branch and cut" algorithms to find exact ground states of
Ising spin glasses\cite{rieger} in fairly large systems in a reasonable
amount of CPU time. For repeated simulated annealings of $N$ different
samples
the CPU time becomes prohibitive for $L>8$ in $2d$ and $L>4$ in $3d$. We
therefore chose sizes $L=2,4,8$ in $2d$ and $L=2,3,4$ in $3d$ and the
results are summarized in Fig.(1) for $2d$ and in Fig.(2) for $3d$.

In $2d$, for small disorder $\alpha <\alpha_{c}\approx 0.37$, $J(L)$
increases more slowly than a power of $L$ and seems to flow to a finite
disorder dependent value $J^{*}(\alpha)$ and the disorder strength $|A(L)|$
does not change with $L$, at least for our small sizes, both of which are
completely consistent with analytic RG calculations\cite{scheidl,tang}. At
larger disorder strength $\alpha >\alpha_{c}$, $|A(L)|$ increases and $J(L)$
decreases as $L$ increases, in agreement with the analytic
theory\cite{scheidl,tang}.
In this range of disorder strengths, the system is probably flowing to the
non-superconducting glass fixed point at $J^{*}=0, A^{*}=\pi/2$. When the
disorder is maximal ($\alpha =1$), $|A(L)|$ remains fixed at $\pi/2$ and
$J(L)\sim L^{\theta}$ with the stiffness exponent $\theta\approx -1/2$, in
agreement with other simulations of the gauge glass in $2d$\cite{fty,mjpg}.
The flows shown in Fig.(1) may be regarded as RG flows in a higher
dimensional parameter space projected on to the ($J(L),|A(L)|$) plane, so
the crossing of the two trajectories for $\alpha=0.45$ and $\alpha=0.5$ does
not violate the non-crossing rule. For the $2d$ system, the RG flows are in
the three parameter space of $J,|A|$ and the vortex fugacity
$y$\cite{nskl,scheidl,tang}.
From the data of Fig.(1) and assuming that the trends for small $L$ continue
when $L$ is large, one would conclude that the most likely scenario for the
disordered system in $2d$ is, for weak disorder $\alpha <\alpha_{c}$,
$J(L)\rightarrow J^{*}(\alpha)$ and $|A(L)|=|A(1)|$,
so that the coupling constant scales to a finite but disorder dependent
value while the disorder is marginal. For larger disorder $\alpha
>\alpha_{c}$, $J(L)\rightarrow J^{*}=0$ and $|A(L)|\rightarrow \pi/2$ which
is a non-superconducting disordered state. Our results are consistent in all
respects with recent analytical theory\cite{scheidl,tang} and inconsistent
with the re-entrant\cite{rsn,gk}
and complete destruction of superconductivity\cite{kor} scenarios.
However, they are consistent with a re-entrant scenario in which the
ordered phase extends to $T=0$ for a range of $\alpha$\cite{vagov}. The
flows
for $2d$ of Fig.(1) are consistent with a discontinuous jump in $J^{*}$ at
$\alpha_{c}$ as predicted analytically\cite{scheidl}.

\begin{figure}
\centering
\begin{minipage}{8.0cm}
\epsfxsize= 8 cm \epsfbox{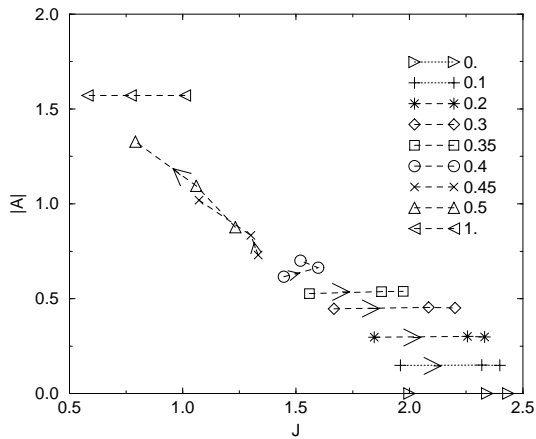}
\end{minipage}
\caption{DWRG flows for the $2d$ superconducting glass
model in the coupling constant-disorder strength ($J,|A|$) plane.
The initial value of $J$ is $J(1)=1$ and the initial disorder strength
$\alpha$ is indicated near corresponding symbols.}
\end{figure}

The results of the simulations of the random gauge model in $3d$ are shown
in Fig.(2). The system sizes are very small ($L=2,3,4$) because of the CPU
time needed to get close to the minimum energies so that irrelevant
variables  are giving large corrections to scaling.
Nevertheless, some qualitative features are
apparent, assuming that the small $L$ trends continue. For small disorder
$\alpha <\alpha_{c}\approx 0.55$, $J(L)\sim L^{d-2}$ as expected and the
disorder strength seems to decrease. It is impossible to say if
$|A(L)|\rightarrow 0$ as one expects but the data is consistent with this.
One is tempted to conclude that, in this regime of weak disorder, the DWRG
flows are to a stable fixed point at $J^{*}=\infty$, $A^{*}=0$ corresponding
to a true superconducting phase. For larger disorder
$\alpha_{c}<\alpha\leq\pi/2$, the disorder increases with $L$ and seems to
flow to its maximum value of $\pi/2$. The coupling $J(L)$ seems to flow to a
finite value which corresponds to a stiffness exponent $\theta =0$. Although
this is consistent with other simulations on the $3d$ gauge
glass\cite{mjpg,rtyf},
our use of the phase representation of eq.(\ref{1}) in the simulations
together with the very small sizes may introduce large corrections to
scaling. Nevertheless, the data is
consistent with a superconducting glass phase which survives at finite $T$.
These considerations suggest that the phase diagram for the model in $3d$ is
similar to that of the corresponding infinite range model\cite{sh}.

\begin{figure}
\centering
\begin{minipage}{8.0cm}
\epsfxsize= 8 cm \epsfbox{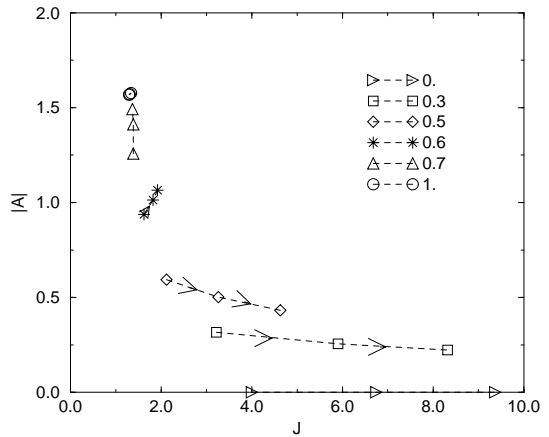}
\end{minipage}
\caption{DWRG flows for the model in $3d$.}
\end{figure}

Our conclusions from our new $T=0$ DWRG which follows the flows in two
parameter space are in $2d$, the recent analytic theory
which predicts a quasi long range ordered state for $T<T_{c}(\alpha)$ is the
correct scenario and earlier suggestions of a re-entrant transition to
a disordered phase at low $T$ or no superconductivity at any finite disorder
are ruled out. In $3d$, weak disorder has little or no effect on the
supeconducting phase and there is a critical disorder strength parametrized
by $\alpha=\alpha_{c}$, above which the system is a superconducting glass at
low $T$.

This work was supported by the NSF under Grant No. DMR-9222812. Computations
were performed on the Cray EL98 at the Theoretical Physics Computing
Facility at Brown University. JMK thanks B. Grossman and A. Vallat for
invaluable discussions about gauge glasses. MVS is grateful to M. Cieplak,
M.J.P. Gingras and A.V. Vagov for correspondence.

\end{multicols}
\end{document}